\documentclass[aps,preprint,amsmath,amssymb]{revtex4}
\usepackage{graphicx,color}

\def\be{\begin{eqnarray}}
\def\ed{\end{eqnarray}}
\def\non{\nonumber}

\begin{document}

\title{\Large \bf New charged Higgs effects on $\Gamma_{K_{e2}}/\Gamma_{K_{\mu 2}}$, $f_{D_s}$ and  ${\cal B}(B^+\to \tau^+ \nu)$ in the Two-Higgs-Doublet model}

\author{ \bf Y.~H.~Ahn$^1$\footnote{E-mail:
        yhahn@phys.sinica.edu.tw}, Chuan-Hung Chen$^{2,3}$\footnote{Email:
physchen@mail.ncku.edu.tw} }

\affiliation{$^{1}$Institute of Physics, Academia Sinica, Taipei 115, Taiwan \\
$^{2}$Department of Physics, National Cheng-Kung
University, Tainan 701, Taiwan \\
$^{3}$National Center for Theoretical Sciences, Hsinchu 300, Taiwan
 }

\date{\today}

\begin{abstract}
A new scheme for Yukawa matrices is proposed for free of flavor changing neutral currents (FCNCs) at tree level  in general two-Higgs-doublet model (THDM) without imposing symmetry. We find that the new type couplings of charged Higgs to fermions not only depend on the flavors, but also can be ascribed by new CP violating phases. Unlike conventional THDM, the resulted new charged Higgs effects could have the specialties: (a) the influence on $\Gamma(K^\pm_{e 2})/\Gamma(K^\pm_{\mu 2})$ cannot be eliminated, (b) the decay constant of $D_s$ could be enhanced, and (c) enhancement of  branching ratio for $B^+\to \tau^+ \nu$ could be achieved.

\end{abstract}

\maketitle %


Despite most experiments in agreement with the standard model (SM)
predictions, it is believed that SM is an effective theory at electroweak scale.
For understanding the neutrino oscillations, matter-antimatter asymmetry,
dark matter, etc., new physics beyond the SM should be included. A direct way to explore the exotic events is through high energy collisions, such as Large Hadron Collider (LHC) and International Linear Collider (ILC). However, by precision measurement and small theoretical uncertainties, low energy system could also provide a good environment to detect the new physics effects indirectly.

Recently, since the experimental data have been reached a new precise level and some inconsistent results between theory and experiment are revealed,  the potential of testing the SM in $P\to \ell \bar\nu$ has been studied broadly and seriously \cite{Masiero:2005wr,Dobrescu:2008er,Kundu:2008ui,Isidori:2006pk,Chen:2006nua,Akeroyd:2007eh}, where $P$ denotes the charged $K$, $D$ and $B$ mesons. The current relevant measurements and SM predictions for the leptonic decays are summarized as follows:  the world average (WA) of $f_{D_s}$ extracted by $D_s \to \ell^+ \nu$ is \cite{Rosner:2010ak}
 \be
 f_{D_s}&=& 257.5 \pm 6.1\ \texttt{MeV} \,,
 \ed
where the new measurement on $D_s\to \tau^+ \nu$ by BaBar Collaboration \cite{:2010qj} has been taken into account; however, the theoretical average of HPQCD ($241 \pm 3$ MeV) \cite{Follana:2007uv}  and Fermilab/MILC ($260\pm 10$ MeV) \cite{Bazavov:2009ii} is given by $f^{\rm Latt}_{D_s} = 242.6 \pm 2.9$ MeV \cite{Kronfeld:2009cf}. Other theoretical prediction could be referred to Ref.~\cite{Gershtein:1976aq}.  The $2.4\sigma$ deviation from the measurement is the so-called $f_{D_s}$ puzzle. The WA of BR for $B \to \tau \nu$ now is \cite{Eigen:2009qg}
 \be
 {\cal B}(B^+\to \tau^+ \nu)&=& (1.73\pm 0.37) \times 10^{-4}\,.
 \ed
With $|V_{ub}|=(0.393\pm 0.036) \%$ \cite{PDG08} and average of $f_B$ in calculations of HPQCD ($190 \pm 13$ MeV) \cite{Gamiz:2009ku} and Fermilab/MILC ($195\pm 11$ MeV) \cite{Bernard:2009wr}, the SM prediction can be read by  ${\cal B}^{\rm SM}(B^+\to \tau^+ \nu)=(1.01\pm 0.19 \pm 0.13) \times 10^{-4}$. The theoretical prediction is somewhat smaller than experimental value.
As to $K^+\to \ell^+ \nu$ decays, the WA with the measurements of KLOE \cite{:2009rv} and NA62 \cite{Goudzovski:2009me} is
 \be
R_K &=& \frac{\Gamma_{K_{e 2}}}{\Gamma_{K_{\mu 2}}}=\frac{\Gamma(K^+\to e^+ \nu)}{\Gamma(K^+\to \mu^+ \nu)}\,, \non \\
&=& (2.498\pm 0.014)\times 10^{-5}\,,
 \ed
whereas the SM predicts $R^{\rm SM}_{K}=(2.477 \pm 0.001)\times 10^{-5}$ \cite{Cirigliano:2007xi}. By the accurate measurement of $0.4\%$ on $R_K$ \cite{Goudzovski:2009me}, the violation of lepton universality may have the chance to be explored in mesonic decays.


To investigate the impact of new physics effects on $P^+\to \ell^+ \nu$, in this paper, we concentrate on the charged Higgs mediated effects in the two-Higgs-doublet model (THDM) \cite{Lee:1973iz}. It is well known that the general THDM  leads to flavor changing neutral currents (FCNCs) at the tree level. For avoiding the large tree level FCNCs, some discrete or global $U(1)$ symmetry has to be imposed so that one Higgs doublet couples to up-type quarks while another one couples to down-type quarks \cite{Glashow:1976nt}, where now it is named by type-II THDM. Although in large $\tan\beta$ scenario (definition given below), the charged Higgs has an important impact on the $P^+ \to \ell^+ \nu$, however its contributions are destructive and make the theoretical results further depart from experiments \cite{Hou:1992sy,Chen:2006nua}. In addition, the nonuniversal lepton couplings in $R_K$ are eliminated, i.e. the value of $R_K$ in type-II THDM is the same as that in the SM \cite{Masiero:2005wr}. Hence, in order to enhance $f_{D_s}$ and ${\cal B}(B^+\to \tau^+ \nu)$ and show the violation of lepton universality in $R_K$, our purpose is to explore the intriguing scenario for Yukawa matrices which could lead to free of FCNCs at the tree level in the THDM without imposing symmetry.


We start with writing the Yukawa sector in THDM as
 \be
-{\cal L}_Y &=& \bar Q_L Y^U_1 U_R \tilde H_1 + \bar Q_L Y^{U}_2 U_R
\tilde H_2  \non
\\
&+& \bar Q_L Y^D_1 D_R H_1 + \bar Q_L Y^{D}_2 D_R H_2+ h.c.
\label{eq:Yu}
 \ed
with $\tilde H_k=i\tau_2 H^*_k$. We can recombine $H_1$ and $H_2$ so that only one of Higgs doublets develops vacuum expectation value (VEV). Accordingly, the new doublets are expressed by
 \be
 h &=&\sin\beta H_1 + \cos\beta H_2 = \left(
            \begin{array}{c}
              G^+ \\
              (v+h^0 +i G^0)/\sqrt{2} \\
            \end{array}
          \right) \,, \non \\
 H &=& \cos\beta H_{1} - \sin\beta H_2=\left(
            \begin{array}{c}
              H^+ \\
              H^0 +i A^0 \\
            \end{array}
          \right)
 \ed
where $\sin\beta=v_1/v$, $\cos\beta=v_2/v$, $v=\sqrt{v^2_1 + v^2_2}$, $<H>=0$ and $<h>=v/\sqrt{2}$. As a result, Eq.~(\ref{eq:Yu}) can be rewritten by
 \be
 -{\cal L}_{Y} &=& \bar Q_L \bar Y^U_1 U_R \tilde h + \bar Q_L \bar Y^U_2 U_R \tilde H \non\\
 &+& \bar Q_L \bar Y^D_2 D_R h - \bar Q_L \bar Y^D_1 D_R H
 \label{eq:Yu2}
 \ed
with
 \be
 \bar Y^{U(D)}_{1(2)} &=& \sin\beta Y^{U(D)}_1 + \cos\beta Y^{U(D)}_2 \,, \non\\
 \bar Y^{U}_{2} &=& \cos\beta Y^{U}_1 - \sin\beta
 Y^{U}_2\,,\non\\
 \bar Y^{D}_{1} &=& -\cos\beta Y^{D}_1 + \sin\beta
 Y^{D}_2\,.
 \ed
Here, $\bar Y^{U(D)}_{1(2)}$ dictate the masses of quarks while $\bar Y^{U(D)}_{2(1)}$ provide the couplings of new neutral and charged Higgses to the SM particles. We note that the same expression could be applied to leptons and the corresponding Yukawa matrices could be read by $\bar Y^{\ell}_{2,1}$, respectively. Hence, from Eq.~(\ref{eq:Yu2}) the diagonalized mass matrix for fermions is given by
 \be
 m^{\rm dia}_F &=& \frac{v}{\sqrt{2}} V^F_L \bar Y^F_{\alpha}
 V^{F\dagger}_{R}\label{eq:mF}
%
 \ed
where $\alpha=1(2)$ for $F=U(D, \ell)$. Clearly, if $\bar Y^{F}_{1(2)}$ and $\bar Y^{F}_{2(1)}$ cannot be diagonalized simultaneously, the FCNCs at tree level will be induced and associated with doublet $H$. Now our purpose is to look for the nontrivial $\bar Y^{F}_{2(1)}$ so that FCNCs can be avoided. The most obvious solution to the question is the aligned Yukawa matrices, i.e. $\bar Y^{F}_{2(1)} \propto \bar Y^{F}_{1(2)} $ \cite{Pich:2009sp}. Due to the coupling of charged Higgs and charged lepton being proportional to the mass of lepton, this scenario will lead to the ratio, defined by
 \be
 R_{P}&=& \frac{\Gamma(P^{\pm}_{\ell' 2})}{\Gamma(P^{\pm}_{\ell
 2})}\,, \label{eq:RP}
 \ed
to be the same as SM; in other words, the violation of lepton universality in $R_P$ will be canceled. We will show that in some interesting scenario, not only can $\bar Y^{F}_{2(1)}$ and $\bar Y^{F}_{1(2)}$  be diagonalized simultaneously but also the violation of lepton universality could be generated in $R_P$ by $H^{\pm}$-mediated effects.

\begin{figure}[hptb]
\includegraphics*[width=4.5 in]{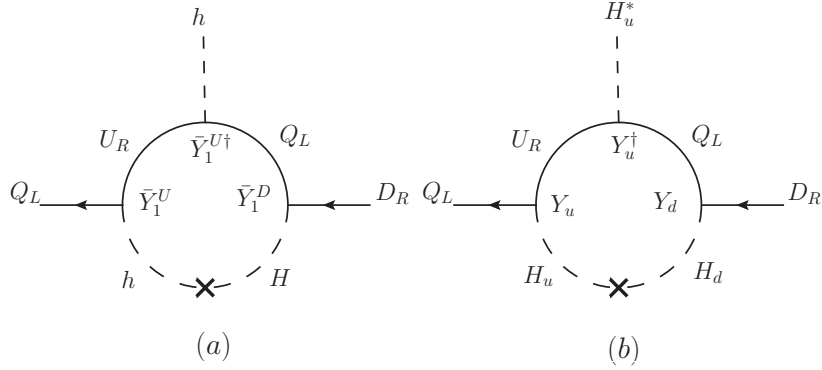}
\caption{ FCNCs induced by radiative corrections (a) without imposing symmetry and (b) with $Z_2$ or $U(1)$ symmetry which is broken softly in Higgs potential.
}
\label{fig:loop}
\end{figure}

To find the suitable $\bar Y^{F}_{2(1)}$ for satisfying our criterions, we set the relevant matrices to be
 \be
 I_{00}&=& \left(
         \begin{array}{ccc}
           a & 0 & 0 \\
           0 & b & 0 \\
           0 & 0 & c \\
         \end{array}\right)\,, \
 I_{12}=\left(
         \begin{array}{ccc}
           0 & a & 0 \\
           b & 0 & 0 \\
           0 & 0 & c \\
         \end{array}
       \right)\,, \non\\
 I_{23}&=& \left(
         \begin{array}{ccc}
           a & 0 & 0 \\
           0 & 0 & b \\
           0 & c & 0 \\
         \end{array}\right)\,, \
 I_{31}= \left(
         \begin{array}{ccc}
           0 & 0 & a \\
           0 & b & 0 \\
           c & 0 & 0 \\
         \end{array}
       \right) \label{eq:Imatrx}
 \ed
with $a$, $b$ and $c$ being arbitrary complex numbers. Multiplying the mass matrix of  Eq.~(\ref{eq:mF}) by $I_{ij}$ following $I_{ij} m^{\rm dia}_{F} I^{\dagger}_{ij}$, $I_{ij} m^{\rm dia}_{F} I_{ij}$ and $I_{ij} m^{\rm dia}_{F} I^{T}_{ij}$, one can find that the resulted new matrices are still diagonal. For illustration, we explicitly express $(\bar m^{\rm dia}_F)_{ij}\equiv I_{ij} m^{\rm dia}_{F}I^{T}_{ij}$ as
\begin{widetext}
 \be
 (\bar m^{\rm dia}_F)_{00} &=&\left(
         \begin{array}{ccc}
           a^2 m_{f1} & 0 & 0 \\
           0 & b^2 m_{f2} & 0 \\
           0 & 0 & c^2 m_{f3} \\
         \end{array}\right)\,, \
 (\bar m^{\rm dia}_F)_{12} = \left(
         \begin{array}{ccc}
           a^2 m_{f2} & 0 & 0 \\
           0 & b^2 m_{f1} & 0 \\
           0 & 0 & c^2 m_{f3} \\
         \end{array}\right)\,, \non \\
  (\bar m^{\rm dia}_F)_{23} &=&\left(
         \begin{array}{ccc}
           a^2 m_{f1} & 0 & 0 \\
           0 & b^2 m_{f3} & 0 \\
           0 & 0 & c^2 m_{f2} \\
         \end{array}\right)\,, \
  (\bar m^{\rm dia}_F)_{31} = \left(
         \begin{array}{ccc}
           a^2 m_{f3} & 0 & 0 \\
           0 & b^2 m_{f2} & 0 \\
           0 & 0 & c^2 m_{f1} \\
         \end{array}\right)\,. \label{eq:mFij}
 \ed
\end{widetext}
We find that besides the diagonal form is obtained, the new matrices may not have the mass hierarchy as shown in Eq. (\ref{eq:mF}). Moreover, by the multiplications
of $I_{ij}\times I_{mn}$, more possible patterns can be found. Consequently, a nontrivial and interesting relation between $Y^{F}_{1(2)}$ and $Y^{F}_{2(1)}$ indeed exists and FCNC free at the tree level can be realized in the THDM without imposing symmetry. In order to give a general expression, we formulate the new diagonal matrix as
 \be
\bar m^{\rm dia}_F \equiv I_{\rho\sigma} m^{\rm dia}_{F} \tilde
I_{\rho\sigma} = V^F_L \bar I^{F}_{L\rho\sigma} \frac{v}{\sqrt{2}}
\bar Y^F_{\alpha} \tilde {\bar I}^{F}_{R\rho\sigma} V^{F
\dagger}_{R}\,,
 \ed
where $I_{\rho\sigma}$ could be any one of the matrices shown in Eq.~(\ref{eq:Imatrx}) or their combinations, $\tilde I_{\rho\sigma}$ could be $I^{\dagger}_{\rho\sigma}$, or $I_{\rho\sigma}$ or $I^T_{\rho\sigma}$, $\bar I^{F}_{\chi\rho\sigma}= V^{F\dagger}_{\chi} I_{\rho\sigma}
V^{F}_{\chi}$ with $\chi=L(R)$ being the helicity projection operator. Hence, if we set $\bar Y^{F}_{2(1)} = \bar I^{F}_{L\rho\sigma} \bar Y^{F}_{1(2)} \tilde{\bar I}^{F}_{R\rho\sigma}$, our purpose to find the solution to diagonalizing $\bar Y^{F}_{1(2)}$ and $\bar Y^{F}_{2(1)}$ simultaneously has been achieved. It is worth mentioning that although there are no FCNCs at the tree level, however, due to no symmetry protection, the FCNCs could be induced by radiative corrections, sketched in Fig.~\ref{fig:loop}(a). Nevertheless, due to the soft $Z_2$ or $U(1)$ breaking term, the similar radiative corrections also occur in the type-II THDM, illustrated in Fig.~\ref{fig:loop}(b). Although the loop-suppressed FCNCs could have interesting impacts on rare decays \cite{Chen:2006nua,Isidori:2001fv}, here we only pay attention to the leading effects on tree processes.

We now move forward to the charged Higgs interactions with fermions. Although the elements in Eq.~(\ref{eq:mFij}) do not show the regular hierarchy in masses of fermions, however, due to $a$, $b$ and $c$ being arbitrary complex numbers, we can reparameterize $\bar m^{\rm dia}_{F}$ to be
 \be
\bar m^{\rm dia}_{F}= {\boldsymbol \eta_F} m^{\rm dia}_{F}
 \ed
where ${\boldsymbol \eta_F}={\rm diag}(\eta_{F1}, \eta_{F2}, \eta_{F3})$ is a new diagonal matrix and the elements are undetermined complex parameters. Based on Eq.~(\ref{eq:Yu2}), the corresponding charged Higgs interactions could be written as
 \be
{\cal L}_{H^{\pm}}&=& \left[\bar U_R \bar Y^{U\dagger}_2
D_L  + \bar U_L \bar Y^D_1 D_R + \bar \nu_L \bar Y^\ell_1 \ell_R\right] H^+ + h.c. \,, \non \\
&=& \frac{\sqrt{2}}{v} \bar u\left[  m^{\rm dia\dagger }_{U}
{\boldsymbol \eta}_U^{\dagger}V
P_L + V {\boldsymbol \eta}_D m^{\rm dia}_{D} P_R  \right] d\; H^{+} \non \\
&+& \frac{\sqrt{2}}{v} \bar\nu {\boldsymbol \eta}_{\ell} m^{\rm
dia}_{\ell} P_R \ell\; H^+ + h.c.\,, \label{eq:Hpm_int}
  \ed
where all flavor indices are suppressed and $V=V^{U}_{L} V^{D\dagger}_{L}$ is the Cabibbo-Kobayashi-Maskawa (CKM) matrix. For comparison,  if we take $\eta_{U_i}=\cot\beta$ and $\eta_{D_i}=\eta_{\ell_i}=\tan\beta$ ($i$=1-3), Eq.~(\ref{eq:Hpm_int}) can be restored to the type-II THDM.

 According to Eq.~(\ref{eq:Hpm_int}), the effective Hamiltonian for $d_j\to u_i \ell \bar\nu_{\ell}$ mediated by $H^{\pm}$ is found by
 \be
{\cal H}_{H^\pm}&=& - \frac{G_F V_{ij}}{\sqrt{2}}
\frac{\eta^*_{\ell} m_{\ell}}{m^2_{H^\pm}}  \left[ \eta^*_{U_i} m_{U_i}(\bar u_i d_j)_{S-P} \right.\non\\
&+&\left. \eta_{D_j} m_{D_j}(\bar u_i d_j)_{S+P} \right] (\bar \ell
\nu_{\ell})_{S-P}
 \ed
where we have used $G_F=1/\sqrt{2}v^2$ and $(\bar u_i d_j)_{S\pm P}=\bar u_i (1\pm\gamma_5) d_j$. With the definition of $P$-meson decay constant, given by
 \be
 \langle 0 | \bar q' \gamma_{\mu} \gamma_5 q | \bar
 P(p)\rangle &=& -if_P p_{\mu}\,, \non \\
 \langle 0 | \bar q'  \gamma_5 q | \bar
 P(p) \rangle &=&i \frac{f_P m^2_P}{m_q + m_{q'}}\,,\non
 \ed
the $H^{\pm}$-mediated transition matrix element combined with SM
contribution for $\bar P\to \ell \bar \nu_{\ell}$ is obtained as
 \be
{\cal M}^{SM+H^\pm}_{\bar P\to \ell \bar \nu_{\ell}}&=& -i
\frac{G_F}{\sqrt{2}} V_{ij} m_{\ell}f_P r^{\ell}_{P} (\bar \ell
\nu_{\ell})_{S-P} \label{eq:amp}
 \ed
 with
 \be
 r^{\ell}_P= 1+ \frac{\eta^*_{\ell} m^2_P}{m^2_{H^\pm} }
\frac{\eta^*_{U_i} m_{U_i} - \eta_{D_j} m_{D_j} }{m_{U_i}+
m_{D_j}}\,. \label{eq:rpl}
 \ed
We learn that the $H^\pm$-mediated contribution is only associated with the factor $r^{\ell}_{P}$ and it depends on the species of lepton due to the appearance of $\eta_{\ell}$.
Since $\eta_{U_i}$, $\eta_{D_j}$ and  $\eta_{\ell}$ are all free parameters, in order to make the results be more predictive, we can adopt a
simple scenario. As mentioned earlier, $\eta_{U_i}$ and $\eta_{D_i, \ell}$ play the role of $\cot\beta$ and $\tan\beta$ in the type-II THDM, respectively. If the new $H^{\pm}$-mediated effects would like to satisfy the constraints of current data such as $b\to s\gamma$, it is plausible to set $|\eta_{U_i}| \ll |\eta_{D_i}|\approx |\eta_{\ell}|$ \cite{Deschamps:2009rh}. As a consequence, $r^{\ell}_{P}$ could be simplified by
 \be
r^{\ell}_{P} \approx 1- \frac{ m^2_P}{m^2_{H^\pm} } \frac{  m_{D_j}
}{m_{U_i}+ m_{D_j}}|\eta_{D_j}|^2 e^{i \phi^{\ell}_{D_j}}  \,. \label{eq:r_P}
 \ed
Intriguingly, in this plain scenario we see that the dependence of lepton flavor in $r^{\ell}_{P}$ can be ascribed to the phase factor $\phi^{\ell}_{D_j}$.  Since $\phi^{\ell}_{D_j}$ are the new physical phases, in general, they cannot be rotated away. If we enforce $\eta_{D_i}=\tan\beta$, we see that the magnitude of charged Higgs effects is the same as that in  type-II THDM. In other words, apart from the new phase factor $\phi^{\ell}_{D_j}$, we do not  introduce a new enhanced factor.

\begin{figure}[hptb]
\includegraphics*[width=3. in]{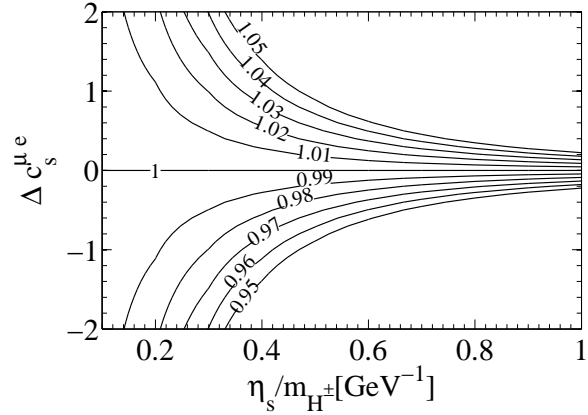}
\caption{Contour for $\Gamma(K^\pm_{e 2})/\Gamma(K^\pm_{\mu 2})$ as
a function of $\eta_s /m_{H^\pm}$ and $\Delta c^{\mu e}_s$, where the
values in the figure are $R^{\rm Exp}_{K}/R^{\rm SM}_{K}$.}
 \label{fig:RK}
\end{figure}

In order to display the new physics effects numerically, we investigate the influence of charged Higgs on $R_K$ for $K_{\ell 2}$, on $f_{D_s} r^\ell_{D_s}$ for $D_s\to \ell^+  \nu_{\ell}$ decays and on BR for $B\to \ell^+ \nu_{\ell}$, respectively. Using Eqs.~(\ref{eq:RP}) and (\ref{eq:amp}), the ratio of $\Gamma(K^\pm_{e2})$ to $\Gamma(K^\pm_{\mu2})$ can be expressed by
 \be
R_K=
R^{\rm SM}_K\left( 1 + \frac{ m^2_K}{m^2_{H^\pm} } \frac{ m_{s}
}{m_{u}+ m_{s}} \eta^2_{s} \Delta c^{\mu e}_s \right)
 \ed
with $\Delta c^{\mu e}_{s} = \cos(\phi^{\mu}_{s})-\cos(\phi^{e}_{s})$, where because of the second term in the brackets being much smaller than unity, we have neglected the terms whose the order is higher than $m^2_K \eta^2_s/m^2_{H^\pm}$. The resulted numerical values as a function of $\eta_s/m_{H^\pm}$ and $\Delta c^{\mu e}_s$ are presented in Fig.~\ref{fig:RK}. The values in the figure denote the ratio $R^{\rm Exp}_{K}/R^{\rm SM}_{K}$. We see clearly that due to the lepton flavor dependent phases, $H^\pm$-mediated contributions could modify the SM prediction and be still consistent with current data.

Similarly, the same discussions could be applied to $R_{\pi}=\Gamma_{\pi_{e 2}}/\Gamma_{\pi_{\mu 2}}$, where the current world average is $R_{\pi}=(1.230 \pm 0.004)\times 10^{-4}$  \cite{PDG08} while the SM prediction is $R^{\rm SM}_{\pi}=(1.2352 \pm 0.0001)\times 10^{-4}$ \cite{Cirigliano:2007xi}. By the results, it seems that  $R_{\pi}$ could give a more strict constraint on free parameters. For clarifying this point, we use the result of Eq. (\ref{eq:r_P}) and choose some typical values of parameters as the illustration. With $m_u\sim m_d$, $m_\pi = 0.14$ GeV, $m_{H^\pm}=200$ GeV and $\eta_d=50$, Eq. (\ref{eq:r_P}) could be estimated to be
 \be
 r^{\ell}_{\pi} \approx 1 - 1.2\times 10^{-3} e^{i\phi^{\ell}_{d}}\,.
 \ed
As a result, the charged Higgs effect on $R_{\pi}$ is of order of $10^{-3}$, which is smaller than that on  $R_{K}$ by a factor $m^2_K/m^2_\pi \approx 12$. Hence, $R_K$ is more sensitive to the charged Higgs effects.

\begin{figure}[hptb]
\includegraphics*[width=3. in]{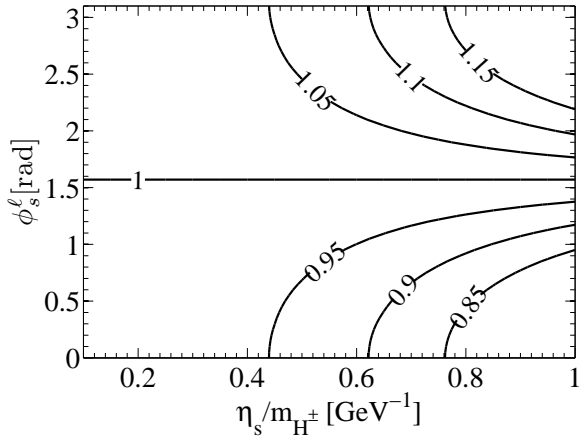}
\caption{Contour for $f_{D_s}$ as a function of $\eta_s/m_{H^\pm}$
and new phase $\phi^{\ell}_{s}$, where the values in the figure
denote $f^{\rm Exp}_{D_s}/f^{\rm Latt}_{D_s}$.}
 \label{fig:rDs}
\end{figure}

Although $P\to \ell^+ \nu \gamma$ will contribute to the measurement of $P\to \ell^+ \nu$ and contaminate the extraction of $P$-meson decay constant, however, it has been studied that the radiative corrections to $D^+_s\to \mu^+(\tau^+ ) \nu$ are around (below) $1\%$ \cite{Dobrescu:2008er,Burdman:1994ip}. Consequently, if we regard the corresponding CKM matrix element as a certain input, the decay constant of charmed meson could be taken as the physical quantity to test the SM. For $D^+_{s,d}\to \ell^+ \nu$ decays, since the $H^\pm$-mediated effects are proportional to the masses of down type quarks, from Eq.~(\ref{eq:amp}) one can understand that Lattice calculations and data have consistent results in $f_{D_d} $; however due to $r^{\ell}_{D_s}$ being not negligible, a sizable difference in $f_{D_s} $ between Lattice \cite{Follana:2007uv,Bazavov:2009ii} and data can occur. Hence, for displaying the $H^{\pm}$ effects, the relationship in decay constant of  $D_s$ between observed value and Lattice can be formulated by
 \be
f^{\rm Exp}_{D_s}\approx f^{\rm Latt}_{D_s} \left|
1-\frac{m^2_{D_s}}{m^2_{H^\pm}} \frac{m_s}{m_s + m_c} \eta^2_s
\cos\phi^{\ell}_{s}\right|\,,
 \ed
where we have assumed that $V_{us}$ is known and its uncertainty does not have a significant effect on the decay constant of $D_s$.
Taking $m_c(2\rm GeV)=1.27$ GeV, $m_s(2\rm GeV)=0.104$ GeV and $m_{D_s}=1.968$ GeV \cite{PDG08}, the numerical values of $f^{\rm Exp}_{D_s}/f^{\rm Latt}_{D_s}$ as a function of $\eta_s/m_{H^\pm}$ and $\phi^{\ell}_{s}$ are plotted in Fig.~\ref{fig:rDs}. Owing to the appearance of $\cos\phi^\ell_s$, clearly the sign of $H^{\pm}$ contribution could be flipped and the $f_{D_s}$ puzzle is solved in the general THDM.

\begin{figure}[hptb]
\includegraphics*[width=3. in]{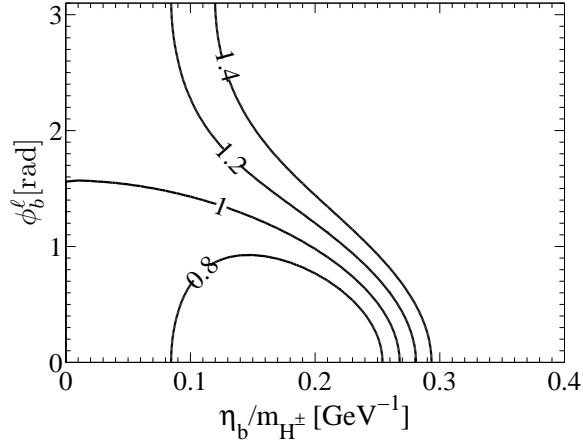}
\caption{Contour for BR of $B^+\to \ell^+ \nu$ as a
function of $\eta_b/m_{H^\pm}$ and new phase $\phi^{\ell}_{b}$,
where the values in the figure denote ${\cal B}^{\rm Exp}(B^+\to
\ell^+ \nu)/{\cal B}^{\rm SM}(B^+\to \ell^+ \nu)$.}
 \label{fig:rBu}
\end{figure}

According to Eq.~(\ref{eq:amp}),  the BR for $B^+\to \ell^+ \nu$ can be straightforward written by
 \be
{\cal B}(B^+\to \ell^+ \nu)&=&{\cal B}^{\rm SM}(B^+ \to \ell^+
\nu)|r^{\ell}_{B}|^2
 \ed
 with
 \be
r^{\ell}_{B}=1-\left(\frac{\eta_b \, m_B}{m_{H^\pm}}\right)^2
e^{i\phi^{\ell}_{b}}  \,.
 \ed
Due to $B$ being a heavy meson, unlike previous cases, we cannot neglect the associated higher order terms. Accordingly, the contour for the influence of $H^\pm$ on $B^+\to \ell^+ \nu$ as a function of $\eta_b/m_{H^\pm}$ and $\phi^{\ell}_{b}$ is plotted in Fig.~\ref{fig:rBu}, where the values appeared in the figure represent the ratio of ${\cal B}^{\rm Exp}(B^+\to \ell^+ \nu)/{\cal B}^{\rm SM}(B^+\to \ell^+ \nu)$. We see clearly that the BR for $B^+\to \tau^+ \nu$ can be enhanced to the value of world average.

\begin{figure}[hptb]
\includegraphics*[width=3. in]{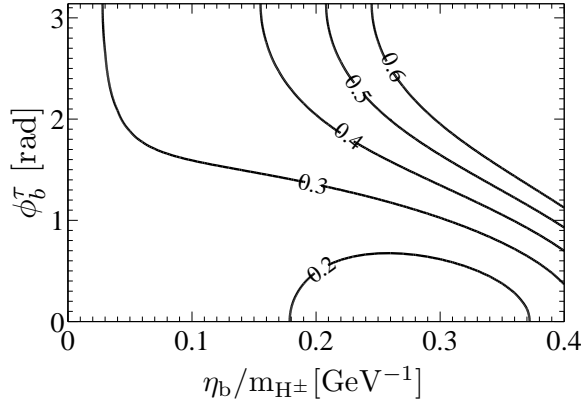}
\caption{Contour for the ratio ${\cal B}(B^-\to D \tau \bar\nu_\tau )/{\cal B}(B^-\to D \ell' \bar\nu_{\ell'})$ as a function of $\eta_b/m_{H^\pm}$ and new phase $\phi^{\tau}_{b}$.}
 \label{fig:rBD}
\end{figure}

In addition to the leptonic decays, we could also study the charged Higgs on semileptonic B decays, such as $B\to (P, V) \ell \nu$. Since the contributions of charged Higgs to light lepton are helicity suppression, thus we just focus on tauon related processes. Recently, BaBar \cite{:2009xy} and Belle \cite{Adachi:2009qg} collaborations have observed $B^-\to D \tau \bar \nu_\tau$ decays to be
 \be
 R_{\bar B\to D \tau \bar\nu_\tau}&=& \frac{{\cal B}(B^-\to D^0 \tau \bar\nu_\tau)}{{\cal B}(B^-\to D^0 \ell' \bar\nu_{\ell'})}=\left\{
                               \begin{array}{c}
                                 (41.6\pm 11.7\pm 5.2)\% \ \  \texttt{ \rm BABAR \cite{:2009xy}}\,, \\
                                 (48^{+22+6}_{-19-5})\% \hspace{1.7 cm} \ \texttt{ \rm BELLE \cite{Adachi:2009qg}}\,,\\
                               \end{array}
                             \right.
  \ed
where $\ell'=e,\,\mu$ and the SM prediction is $R^{\rm SM}_{\bar B\to D \tau \bar\nu_\tau}\approx 0.30$ \cite{Chen:2006nua}. Although the errors of current data are still large, however, it will be a strong hint of new physics if any significant deviation from the SM prediction is found in the future B-factory. For dealing with the decay for $B^-\to D \tau \bar\nu_\tau$, the transition matrix element by SM and $H^\pm$ contributions is written by
 \be
M( \bar B\to D \tau \bar \nu_\tau)&=&\langle  \tau \bar\nu_\tau D| H_{\rm eff}| \bar B\rangle =
\frac{G_F V_{cb}}{\sqrt{2}} \left[
 \langle D| \bar c \gamma_{\mu}(1-\gamma_5)b|  \bar B\rangle \bar\tau\gamma^{\mu} (1-\gamma_5) \nu_{\tau}
 \right. \nonumber \\
 && \left. -\delta_{H}
 \langle D| \bar c (1+\gamma_5)b| \bar B\rangle \bar \tau (1-\gamma_5) \nu_{\tau}
\right]  \label{eq:amp_p}
 \ed
with $\delta_H= m_b m_\tau \eta^2_b/m^2_{H^\pm} e^{i\phi^\tau_{b}}$.
To get the hadronic QCD effects,
we parametrize the $\bar B\to D$ transition as \cite{Chen:2006nua}
 \be
\langle D(p_{D}) | \bar c \gamma^{\mu}  b|  \bar B(p_B)\rangle
&=&
f^{BD}_{+}(q^2)\left(P^{\mu}-\frac{P\cdot q}{q^2}q^{\mu}
\right)+f^{BD}_{0}(q^2) \frac{P\cdot q}{q^2} q_{\mu}\,,
 \label{eq:bpff}
 \ed
with $P=p_{B}+p_{D}$ and $q=p_{B}-p_{D}$. Since the scalar form factor associated with $H^\pm$ contributions is unknown, therefore, for calculating the ratio, we adopt the parametrization given by \cite{Deschamps:2009rh}
 \be
R_{\bar B\to D \tau\nu_\tau}=\frac{{\cal B}(B^- \to D \tau \bar\nu_\tau )}{{\cal B}(B^-\to D \ell' \bar\nu_{\ell'})}= 0.2970 + 0.1065Re(s_H)+0.0178 |s_H|^2
 \ed
with
 \be
 s_H = -\frac{m^2_B -m^2_D}{1-m_c/m_b} \frac{\eta^2_b}{m^2_{H^\pm}} e^{i\phi^\tau_b}\,,
 \ed
where we have neglected the small contributions from light leptons.
Consequently, the contour for $R_{\bar B\to D \tau \bar\nu_\tau}$ as a function of $\eta_b/m_{H^\pm}$ and $\phi^\tau_b$  is displayed in Fig.~\ref{fig:rBD}. We have demonstrated that ${\cal B}(B^- \to D \tau \bar\nu_\tau)$ is also sensitive to the effects of $H^\pm$.

In summary, we find a new scheme for Yukawa couplings in general THDM without imposing symmetry. The scheme not only can avoid FCNC at tree level but also provides a novel couplings of charged Higgs to fermions. With the constraint of $b\to s\gamma$, we find that the violation of lepton universality can be simplified to be the flavor dependent CP violating phase factor, $\phi^{\ell}_{D_j}$. Unlike conventional type-II THDM, the new $H^{\pm}$-mediated effects have the specialties: (a) the influence on $\Gamma(K^\pm_{e 2})/\Gamma(K^\pm_{\mu 2})$ cannot be eliminated, (b) the decay constant of $D_s$ could be enhanced, and (c) enhancement of branching ratio for $B^+\to \tau^+ \nu$ could be accomplished.\\

 \noindent {\bf Acknowledgments}

This work is supported by the National Science Council of R.O.C.
under Grant Nos. NSC-97-2112-M-006-001-MY3 and NSC-97-2112-M-001-004-MY3.

\end{document}